\documentstyle[aps]{revtex}

\begin{document}
\title{Reply on `comment on our paper `Single two-level ion in an
anharmonic-oscillator trap: Time evolution of the Q function and population
inversion ''}
\author{S. Shelly Sharma}
\address{Departamento de F\'{i}sica, Universidade Estadual de Londrina, \\
Londrina,\\
Paran\'{a} 86051-970, Brazil}
\author{N. K. Sharma}
\address{Departamento de Matem\`{a}tica, Universidade Estadual de Londrina, \\
Londrina, \\
Paran\'{a} 86051-970, Brazil}
\author{Larry Zamick}
\address{Department of Physics and Astronomy, Rutgers University, Piscataway\\
New Jersey 08409, USA}
\maketitle

\begin{abstract}
We show here that the model Hamiltonian used in our paper for ion vibrating
in a $q$-analog harmonic oscillator trap and interacting with a classical
single-mode light field is indeed obtained by replacing the usual bosonic
creation and annihilation operators of the harmonic trap model by their $q$%
-deformed counterparts. The approximations made in our paper amount to using
for the ion-laser interaction in a $q$-analog harmonic oscillator trap, the
operator $F_{q}=e^{-(\frac{\left| \epsilon \right| ^{2}}{2})}e^{i\epsilon
A^{\dagger }}e^{i\epsilon A}$ , which is analogous to the corresponding
operator for ion in a harmonic oscillator trap that is $F=e^{-(\frac{\left|
\epsilon \right| ^{2}}{2})}e^{i\epsilon a^{\dagger }}e^{i\epsilon a}$. Here $%
a$ and $a^{\dagger }$ are the usual creation and annihilation operators,
whereas $A$ and $A^{\dagger }$ are the q-deformed bosonic creation and
annihilation operators. There is no problem with diagonalising this operator
using the basis states $\left| g,m\right\rangle $ and $\left|
e,m\right\rangle $ , where $m$ stands for the motional number state. In our
article we do not claim to have diagonalised the operator $%
F_{q}=e^{i\epsilon (A^{\dagger }+A)}$ for which the basis states $\left|
g,m\right\rangle $ and $\left| e,m\right\rangle $ are not analytic vectors.
\end{abstract}


$\Omega \Omega $In our article\cite{shel97} we have used only q-deformed
bosonic creation and annihilation operators, which in the limit $%
q\rightarrow 1$ coincide with the harmonic oscillator creation and
annihilation operators. In this reply, in order to clarify things we shall
make a distinction between the two types of operators and discuss the
ion-laser system Hamiltonian separately for these two cases.

\medskip

\subsection{The Hamiltonian for ion in a harmonic Oscillator trap}

The system that consists of the ion moving in a harmonic oscillator trap
potential and interacting with a classical single-mode light field of
frequency $\omega _{l}$ is described by the Hamiltonian, 
\begin{equation}
H=\frac{1}{2}\hbar \omega (a^{\dagger }a+aa^{\dagger })+\frac{1}{2}\hbar
\triangle \sigma _{z}+\frac{1}{2}\hbar \Omega (F\sigma ^{+}+F^{\dagger
}\sigma ^{-})  \label{eq1}
\end{equation}
where $\Delta =\omega _{a}-\omega _{l}$ , is the detuning parameter and $%
\Omega $ is the Rabi frequency of the system. The operator $F$ stands for%
\cite{bloc92} $\exp (ikx)=\exp [i\epsilon (a^{\dagger }+a)].$ The parameter $%
\epsilon =\sqrt{\frac{E_{r}}{E_{t}}}$ is a function of the ratio of the
recoil energy of the ion $E_{r}=\frac{\hbar ^{2}k^{2}}{2m}$and the
characteristic trap quantum energy $E_{t}=\hbar \omega $. Here $k$ is the
wave vector of the light field and $x$ position quadrature of the center of
mass. The second term in the Hamiltonian refers to the energy associated
with internal degrees of freedom of the ion, whereas the third term is the
interaction of the ion with the light field. For harmonic oscillator trap
one has 
\begin{equation}
\lbrack a,a^{\dagger }]=1\medskip \ ;\medskip \ \ \ Na^{\dagger }-a^{\dagger
}N=a^{\dagger }\medskip \ ;\medskip \ Na-aN=-a.  \label{eq2}
\end{equation}

By using a special form of Baker-Hausdorff Theorem, which reads\cite{Klau}, 
\begin{equation}
e^{X+Y}=e^{-\frac{1}{2}\left[ X,Y\right] }e^{X}e^{Y}  \label{eq3}
\end{equation}
and is valid whenever both $X$ and $Y$ commute with the commutator $Z=\left[
X,Y\right] $, the operator $F$ may be rewritten as a product of operators i.e

\begin{equation}
F=\exp [i\epsilon (a^{\dagger }+a)]=e^{\left( \frac{\left| \epsilon \right|
^{2}}{2}[a^{\dagger },a]\right) }e^{i\epsilon a^{\dagger }}e^{i\epsilon
a}=e^{\left( \frac{-\left| \epsilon \right| ^{2}}{2}\right) }e^{i\epsilon
a^{\dagger }}e^{i\epsilon a}.  \label{eq4}
\end{equation}

Here we examine in some detail as to what kind of processes are implied by
the operator 
\begin{equation}
F=\left[ \exp \left( \frac{-\left| \epsilon \right| ^{2}}{2}\right) \sum_{n}%
\frac{\left( i\epsilon \right) ^{n}a^{\dagger n}}{n!}\sum_{k}\frac{\left(
i\epsilon \right) ^{k}a^{k}}{k!}\right] .  \label{eq5}
\end{equation}

The terms for $n>k$ correspond to the increase in energy by $(n-k)$ quanta
while the one's with $n<k$ can destroy $(k-n)$ quanta thus altering the
amount of energy linked with the center of mass motion. For $(n=k),$ we have
diagonal contributions. The contribution from a particular term with
operators $a^{\dagger n}a^{k}$is determined by the coefficient $\exp \left( 
\frac{-\left| \epsilon \right| ^{2}}{2}\right) \left[ \frac{\left( i\epsilon
\right) ^{n+k}}{n!k!}\right] $.

\subsection{The Hamiltonian for ion in a $q$-analog harmonic oscillator trap}

Next we consider a single two level ion having ionic transition frequency$\
\omega _{a}$ in a quantized $q$-analog quantum harmonic oscillator trap($q$%
-deformed harmonic oscillator trap) interacting with a single mode
travelling light field. The creation and annihilation operators for the trap
quanta satisfy the following q-commutation relations,

\begin{equation}
AA^{\dagger }-qA^{\dagger }A=q^{-N}\medskip \ ;\medskip \ \ \ NA^{\dagger
}-A^{\dagger }N=A^{\dagger }\medskip \ ;\medskip \ NA-AN=-A  \label{eq6}
\end{equation}

Here N is the number operator. The operators $A$ and $A^{\dagger }$ act in a
Hilbert space with basis vectors $\left| n\right\rangle $, $n=0,1,2,...,$
given by,

\begin{equation}
\left| n\right\rangle =\frac{(A^{\dagger })^{n}}{([n]_{q}!)^{\frac{1}{2}}}%
\left| 0\right\rangle  \label{eq7}
\end{equation}
such that $N\left| n\right\rangle =n\left| n\right\rangle .$ The vacuum
state is $A\left| 0\right\rangle =0.$ We define here $[x]_{q}$ as 
\begin{equation}
\lbrack x]_{q}=\frac{q^{x}-q^{-x}}{q-q^{-1}}  \label{eq8}
\end{equation}
and the $q$-analog factorial $[n]_{q}!$ is recursively defined by

[0]$_{q}!=[1]_{q}!=1$ and $[n]_{q}!=[n]_{q}[n-1]_{q}!.$ It is easily
verified that 
\begin{equation}
A^{\dagger }\left| n\right\rangle =[n+1]_{q}^{\frac{1}{2}}\left|
n+1\right\rangle \medskip \ ;\medskip \ A\left| n\right\rangle =[n]_{q}^{%
\frac{1}{2}}\left| n-1\right\rangle  \label{eq9}
\end{equation}
and $N$ is not equal to $A^{+}A.$

\medskip The $q$-deformed operators $A$ and $A^{\dagger }$ are related to $a$
and $a^{\dagger }$ through

\begin{equation}
A=a\,f(N)\qquad ;\qquad A^{\dagger }=f(N)\,a^{\dagger }  \label{eq10}
\end{equation}

where $f(N)=\left( \frac{[N]_{q}}{N}\right) ^{\frac{1}{2}}$ and $%
N=a^{\dagger }a$.

\medskip

The Hamiltonian for an ion interacting with light in a $q$-analog harmonic
oscillator trap may now be written as

\begin{equation}
H_{q}=\frac{1}{2}\hbar \omega (A^{\dagger }A+AA^{\dagger })+\frac{1}{2}\hbar
\triangle \sigma _{z}+\frac{1}{2}\hbar \Omega (F_{q}\sigma
^{+}+F_{q}^{\dagger }\sigma ^{-})  \label{eq11a}
\end{equation}
where by analogy with Eq. (\ref{eq4}) we choose

\begin{equation}
F_{q}=e^{\left( \frac{-\left| \epsilon \right| ^{2}}{2}\right) }e^{i\epsilon
A^{\dagger }}e^{i\epsilon A}.  \label{eq11}
\end{equation}
which in the limit $q\rightarrow 1$ reduces to $F$ and can be expanded as 
\begin{equation}
F_{q}=e^{\left( \frac{-\left| \epsilon \right| ^{2}}{2}\right)
}\sum_{n=0}^{\infty }\frac{(i\epsilon )^{n}A^{\dagger }{}^{n}}{n!}%
\sum_{k=0}^{\infty }\frac{(i\epsilon )^{k}A^{k}}{k!}.  \label{eq12}
\end{equation}
Various terms in the expansion of this operator represent processes which
might result in transitions of the center of mass from a given motional
state, in the q-analog trap, to another, while loosing or gaining energy.
The coefficient of the product of operators $A^{\dagger }{}^{n}A^{k}$ is
again $\exp \left( \frac{-\left| \epsilon \right| ^{2}}{2}\right) \left[ 
\frac{\left( i\epsilon \right) ^{n+k}}{n!k!}\right] $, the same as that in
the corresponding term with operators $a^{\dagger n}a^{k}$ in Eq. (\ref{eq4}%
). Intutively this is the correct way of representing the interaction of the
ion and the laser in a q-analog harmonic oscillator trap. This form of the
operator shows that the energy exchange of the center of mass motion of the
ion occurs as the ion moves up or down in the trap. Since $E_{n+1}-E_{n}$ is 
$n$ dependent, it implies an $n$ dependent entanglement of the center of
mass motion and the internal degrees of freedom of the two level atom. Using
Eq. (\ref{eq10}), we can rewrite Eq. (\ref{eq12}) as 
\begin{equation}
F_{q}=e^{\left( \frac{-\left| \epsilon \right| ^{2}}{2}\right)
}\sum_{n=0}^{\infty }\sum_{k=0}^{\infty }\frac{\left( i\epsilon
\,f(N)a^{\dagger }\right) ^{n}{}\left( i\epsilon \,af(N)\right) ^{k}}{n!k!}.
\label{eq13}
\end{equation}
The matrix elements of this operator are well defined(Eq. 14 of ref.\cite
{shel97})and given for $m\leq n$ by 
\begin{equation}
\left\langle m\right| F_{q}\left| n\right\rangle =\frac{e^{\frac{-\left|
\epsilon \right| ^{2}}{2}}(i\epsilon )^{n-m}[m]_{q}^{\frac{1}{2}}!}{[n]_{q}^{%
\frac{1}{2}}!}\sum\limits_{k=0}^{m}\frac{(\epsilon )^{2k}(-1)^{k}[n]_{q}!}{%
k!(n-m+k)![m-k]_{q}!}  \label{eq14}
\end{equation}

Comparing $F_{q}$ with $F$ , we notice that aside from the factor $e^{\left( 
\frac{-\left| \epsilon \right| ^{2}}{2}\right) },$ the effective lamb Dicke
parameter for the loss and gain of motional state energy in a given
interaction process in a $q$-analog trap is $\epsilon f(N)$, that is it
depends on the number of $q$-oscillator quanta linked with the state at a
given moment. For the initial state of the system considered in our paper
that is 
\begin{equation}
\left| g,\alpha \right\rangle _{q}=\frac{1}{\sqrt{\exp _{q}^{\left| \alpha
\right| ^{2}}}}\sum\limits_{n=0}^{\infty }\frac{\alpha ^{n}}{\sqrt{\left[
n\right] _{q}!}}\left| g,n\right\rangle \text{ ,}  \label{eq15}
\end{equation}
the calculated value of $_{q}\left\langle \alpha \right| f^{2}(N)\left|
\alpha \right\rangle _{q}$ is $1.0004$ , for $\alpha =4$ , $\epsilon =0.05$
and $\tau =0.003$. Of course it increases with increasing $q$. We take this
opportunity to correct a typo in the 5$^{th}$line of Sec. V of our paper 
\cite{shel97} and give again the parameters used for the numerical
calculation. The values are $\epsilon =0.05$, $\overline{\omega }=\frac{%
\omega }{\Omega }=50$ and $\overline{\Delta }=\frac{\Delta }{\Omega }=-50$.
In Eq. (\ref{eq13}), if we approximate $\epsilon f(N)$ by $\epsilon
_{q}=\epsilon \sqrt{_{q}\left\langle \alpha \right| f^{2}(N)\left| \alpha
\right\rangle _{q}}$ which for the case cited above gives $\epsilon
_{q}=0.05001$, we can write the effective ion laser interaction operator as 
\begin{equation}
F_{q}= e^{\left( \frac{\left| \epsilon _{q}\right| ^{2}}{2}-\frac{\left|
\epsilon \right| ^{2}}{2}\right) }e^{\left( -\frac{\left| \epsilon
_{q}\right| ^{2}}{2}\right) }e^{i\epsilon _{q}a^{\dagger }}e^{i\epsilon
_{q}a}=e^{\left( \frac{\left| \epsilon _{q}\right| ^{2}}{2}-\frac{\left|
\epsilon \right| ^{2}}{2}\right) }e^{i\epsilon _{q}(a^{\dagger }+\,\;a)}%
\text{ .}  \label{eq16}
\end{equation}

We agree with the authors of the comment that the operator $(A^{\dagger }+A)$
is not a self adjoint operator and the basis vectors $\left| m\right\rangle $
are not analytic vectors of the operator $\exp \left( i\epsilon (A^{\dagger
}+A)\right) $. In our article\cite{shel97} we start with a form for ion in $%
q $-analog harmonic oscillator trap Hamiltonian that is analogous to the
Hamiltonian of Eq. (\ref{eq1}), but in the third paragraph of Sec. III of
our article we make an approximation and put $\exp [i\epsilon (A^{\dagger
}+A)]=e^{(\frac{\left| \epsilon \right| ^{2}}{2}[A^{\dagger
},A])}e^{i\epsilon A^{\dagger }}e^{i\epsilon A}$. We recall here that the
Baker Hausdorff theorem (Eq.(\ref{eq3})) can not be applied here in case
exact commutation relations for the operators $A$ and $A^{\dagger }$as given
by Eq$.($\ref{eq6}) are used. Only for the case $q=1,$ the operators $%
A^{\dagger }$ and $A$ commute with $[A^{\dagger },A]$. So the approximation
amounts to using $[A^{\dagger },A]=-1$ at a certain level. The same
approximation is used again to approximate $e^{(\frac{\left| \epsilon
\right| ^{2}}{2}[A^{\dagger },A])}$ by $e^{(-\frac{\left| \epsilon \right|
^{2}}{2})}$. For the numerical calculation, the matrix elements of Eq. (\ref
{eq14}) go into the calculation of matrix elements of the type $\left\langle
e,m\right| F_{q}\sigma ^{+}\left| g,n\right\rangle $ . These points have
been mentioned but not discussed in detail in our paper\cite{shel97}. The
final form of the model Hamiltonian as used for numerical calculations is
the one given by Eqs. (\ref{eq11a} and \ref{eq11}) and not the one
containing the operator $\exp ([i\epsilon (A^{\dagger }+A)]$ . It is
relevant to recall here that $e^{\left( \frac{-\left| \epsilon \right| ^{2}}{%
2}\right) }e^{i\epsilon A^{\dagger }}e^{i\epsilon A}$ $\neq e^{\left( \frac{%
\left| \epsilon \right| ^{2}}{2}\right) }e^{i\epsilon A}e^{i\epsilon
A^{\dagger }}$ due to Eq. (\ref{eq6}). We conclude that the numerical
calculations presented in ref. \cite{shel97} are perfectly valid.


\end{document}